\documentstyle[aps,prl,preprint]{revtex}

\draft

\begin{document}
\author{Hong-shi Zong$^{1,2}$, Xiang-song Chen$^{1}$, Fan Wang$^{1}$, 
Chao-hsi Chang$^{2,3}$ and En-guang Zhao$^{2,3}$}
\address {$^{1}$ Department of Physics, Nanjing University, Nanjing 210093, P. R. China} 
\address{$^{2}$ CCAST(World Laboratory), P.O. Box 8730, Beijing 100080, P. R. China}
\address{$^{3}$ Institute of Theoretical Physics, Academia Sinica, P.O. Box 2735, Beijing 100080, P. R. China}

\title{Nonperturbative Aspects of Axial Vector Vertex in the Global Color Symmetry Model}
\maketitle

\begin{abstract}
It is shown how the axial vector current of current quarks is related to that of constituent quarks within the framework of the global color symmetry model. Gluon dressing of the axial vector vertex and the quark self-energy functions is described by the inhomogeneous Bethe-Salpeter equation in the ladder approximation and the Schwinger-Dyson equation in the rainbow approximation, respectively.
\end{abstract}

\bigskip


\bigskip

E-mail: zonghs@chenwang.nju.edu.cn.

\bigskip


Being a dressed particle of sea quarks and gluons, the constiuent quark naturally behaves different from the fundamental current quark of the QCD Lagrangian does, most evidently by a much larger mass. The dynamical or constituent masses appear owing to the spontaneous chiral symmetry breaking in QCD. A realistic and a phenomenological mechanism of chiral symmetry breaking is provided by instantons[1-3] and the global color symmetry model(GCM)[4], respectively. Recently, there are evidences that GCM provides a successful description of various nonperturbative aspect of strong interaction physics and QCD vacuum as well as hadronic phenomena at low energies[4-7]. In Refs.[8-10], it was shown how the vector current of current quarks is related to that of the constituent quarks. In this letter, we follow the approach of Refs.[8-10] and study the axial vector current, which is closely related to the quark spin operator in QCD(note that $\bar{q}\vec{\gamma}\gamma^{5} q=q^{\dagger}\vec{\Sigma} q$).

We consider the Euclidean action of the GCM in an external axial vector field 
${\cal{A}}_{\mu 5}(x)$:
\begin{eqnarray}
&&S_{GCM}[\bar{q},q;{\cal{A}}]\nonumber\\
&&=\int d^{4} x \left\{\bar{q}(x)[\gamma\cdot\partial_{x}+i\gamma_{\nu}\gamma_{5}{\cal{A}}_{\nu 5}(x)]q(x)\right\}+\int d^{4}x d^{4} y\left[\frac{g^2_{s}}{2} j^{a}_{\mu}(x) D^{ab}_{\mu\nu}(x-y)j^{b}_{\nu}(y)\right],
\end{eqnarray}
where $j^{a}_{\mu}(x)=\bar{q}(x)\gamma_{\mu}\frac{\lambda^{a}_{c}}{2}q(x)$  denotes the color octet vector current and $g^2_{s}D^{ab}_{\mu\nu}(x-y)$ is the dressed gluon propagator as input in GCM. For convenience, we will employ the Feynman type gauge $D^{ab}_{\mu\nu}(x-y)=\delta_{\mu\nu}^{ab}D(x-y)$ for the model gluon propagator.

Introducing an auxiliary bilocal field $B^{\theta}(x,y)$ and applying the standard bosonization procedure the generating functional of GCM
\begin{equation}
{\cal{Z}}[{\cal{A}}]=\int{\cal{D}} \bar{q}{\cal{D}}q ~~exp^{-S_{GCM}
[\bar{q},q;{\cal{A}}]}
\end{equation}
can be rewritten in terms of the bilocal fields $B^{\theta}(x,y)$
\begin{equation}
{\cal{Z}}[{\cal{A}}]=\int{\cal{D}} B^{\theta} exp^{-S_{eff}[B^{\theta};{\cal{A}}]}
\end{equation}
with the effective bosonic action
\begin{equation}
S_{eff}[B^{\theta},{\cal{A}}]=-Tr ln {\cal{G}}^{-1}[B^{\theta};{\cal{A}}]+\int d^{4}x d^{4}y\frac{B^{\theta}(x,y) B^{\theta}(y,x)}{2 g^{2}_{s} D(x-y)}
\end{equation}
and the quark operator
\begin{equation}
{\cal{G}}^{-1}[B^{\theta};{\cal{A}}]=[\gamma\cdot\partial_{x}+i\gamma_{\nu}\gamma_{5}{\cal{A}}_{\nu 5}(x)]\delta(x-y)+\Lambda^{\theta} B^{\theta}(x,y).
\end{equation}
The matrices $\Lambda^{\theta}=K^{a}\otimes C^{b}\otimes F^{c}$ is determined by Fierz transformation in color, flavor and Lorentz space of the current current interaction in Eq.(1), and are given by
\begin{equation}
\Lambda^{\theta}=\frac{1}{2}\left\{1_{D},i\gamma_{5},\frac{i}{\sqrt[]{2}}\gamma_{\nu},\frac{i}{\sqrt[]{2}}\gamma_{\nu}\gamma_{5}\right\}\otimes\left\{\frac{1}{\sqrt[]{3}}1_{F},\frac{1}{\sqrt[]{2}}\lambda^{a}_{F}\right\}\otimes\left\{\frac{4}{3} 1_{C},\frac{i}{\sqrt[]{3}}\lambda^{a}_{C}\right\}.
\end{equation}

One might expect that the complete set of the 16 Dirac matrices $\left\{1_{D},i\gamma_{5},\gamma_{5}\gamma_{\mu},\sigma_{\mu\nu}\right\}$ must be employed in its description. However, by limiting the gluon two-point function $g^{2}_{s} D(x-y)$ to diagonal components in Lorentz indices, the tensor $\sigma_{\mu\nu}$ is excluded.

In the mean-field approximation, the fields $B^{\theta}(x,y)$ are substituted simply by their vacuum value $B^{\theta}_{0}(x,y)$, which is defined as
$\frac{\delta{S}_{eff}}{\delta{B}}{\mid}_{B_{0}}=0$ and is given by

\begin{equation}
B^{\theta}_{0}[{\cal{A}}](x,y)=g^{2}_{s}D(x-y)tr_{\gamma C}[\Lambda^{\theta}{\cal{G}}_{0}[{\cal{A}}](x,y)],
\end{equation}
where ${\cal{G}}^{-1}_{0}(x,y)$ denotes the inverse propagator with the self-energy $\Sigma(x,y)=\Lambda^{\theta} B^{\theta}_{0}(x,y)$ in the external background field ${\cal{A}}_{\mu 5}(x)$. Because of the external axial vector field, the quark propagator and self energy are not translationally invariant.

It should be noted that both $B^{\theta}_{0}(x,y)$ and ${\cal{G}}^{-1}_{0}(x,y)$ have an implicit dependence on the external background field ${\cal{A}}_{\mu 5}$. If the external field ${\cal{A}}_{\mu 5}$ is switched off, ${\cal{G}}_{0}[{\cal{A}}]$ goes into the dressed quark propagator $G\equiv{\cal{G}}_{0}[{\cal{A}}_{\mu 5}=0]$, which has the decomposition
\begin{equation}
G^{-1}(p)=i\gamma\cdot p+\Sigma (p)=i\gamma\cdot p A(p^2)+B(p^2)
\end{equation}
with
\begin{equation}
\Sigma(p)=\int d^4 x e^{i p\cdot x}[\Lambda^{\theta} B^{\theta}_{0}(x)]=\frac{4}{3}\int \frac{d^4 q}{(2\pi)^4} g^2_{s}D(p-q)\gamma_{\nu}G(q)\gamma_{\nu}
\end{equation}
where the self energy functiona $A(p^2)$ and $B(p^2)$ are
determined by the rainbow Dyson-schwinger equation(DSE)
\[
[A(p^2)-1]p^2=\frac{8}{3}\int \frac{d^{4}q}{(2\pi)^4}g^2 D(p-q)
\frac{A(q^2)p\cdot q}{q^2A^2(q^2)+B^2(q^2)},
\]
\begin{equation}
B(p^2)=\frac{16}{3}\int \frac{d^{4}q}{(2\pi)^4}g^2 D(p-q)
\frac{B(q^2)}{q^2A^2(q^2)+B^2(q^2)}.
\end{equation}
This dressing comprises the notion of $''$constituent quarks$''$ by providing 
a dynamical mass $M(p^2)=B(p^2)/A(p^2)$, reflecting a vacuum configuration with
dynamically broken chiral symmetry. The order parameter for chiral symmetry breaking is the quark or chiral condensate
\begin{equation}
\langle:\bar{q}q:\rangle_{\mu}=(-)tr_{\gamma C}[G_{0}(x,0)]|_{x=0}=(-)\frac{12}{16\pi^2}\int^{\mu}_{0}ds s
\frac{B(s)}{sA^2(s)+B^2(s)}.
\end{equation}
By definition, this is the quark Green function taken at one point; in momentum space it is closed quark loop. Had the quark propagator only the " slash " term, the trace over the spinor indices understood in this loop would give an identical zero. Therefore, chiral symmetry breaking implies that a massless (or nearly massless) quark develops a non-zero dynamical mass(i.e., a "non-slash" term in the propagator).

In coordinate space the dressed axial vector vertex $\Gamma_{\mu 5}(x,y;z)$ is given as the functional derivative of the inverse quark propagator ${\cal{G}}^{-1}_{0}[{\cal{A}}]$ with respect to the external field ${\cal{A}}_{\mu 5}$:
\begin{equation}
\Gamma_{\mu 5}(y_1,y_2;z)=\left[\frac{\delta {\cal{G}}^{-1}_{0}[{\cal{A}}](y_1,y_2)}{\delta {\cal{A}}_{\mu 5}}\right]_{{\cal{A}}_{\mu 5}=0}
\end{equation}

Taking the functional derivative in Eq.(5) and put it into Eq.(12), we have
\begin{equation}
\Gamma_{\mu 5}(y_1,y_2;z)=i\gamma_{\mu}\gamma_{5}\delta(y_1-y_2)\delta(y_1-z)+\left[\frac{\delta\Sigma[{\cal{A}}](y_1,y_2)}{\delta{\cal{A}}_{\mu 5}(z)}\right]_{{\cal{A}}_{\mu 5}=0}.
\end{equation}
The second term on the right-hand side of Eq.(13) can be determined by employing the stationary condition Eq.(7), which, after reversing Fierz recording, can be cast into
\begin{equation}
\Sigma[{\cal{A}}](y_1,y_2)=\frac{4}{3}g^2_{s} D(y_1,y_2)\gamma_{\nu}{\cal{G}}_{0}[{\cal{A}}](y_1,y_2)\gamma_{\nu}.
\end{equation}
In order to find an expression for $[\delta{\cal{G}}_{0}[{\cal{A}}](y_1,y_2)/\delta{\cal{A}}_{\mu}(z)]_{{\cal{A}}=0}$ in terms of quark propagator 
${\cal{G}}_{0}$, it is obvious that ${\cal{G}}^{-1}_{0}[{\cal{A}}]$ can be expanded in powers of ${\cal{A}}$ as follows
\begin{equation}
{\cal{G}}^{-1}_{0}[{\cal{A}}]={\cal{G}}^{-1}_{0}[{\cal{A}}]\mid_{{\cal{A}}_{\mu 5}=0}+\frac{\delta {\cal{G}}^{-1}_{0}[{\cal{A}}]}{\delta {\cal{A}}}\mid_{{\cal{A}}_{\mu 5}=0}{\cal{A}}+\cdot\cdot\cdot=G^{-1}+{\cal{A}}_{\mu 5}\Gamma_{\mu 5}+\cdot\cdot\cdot,
\end{equation}
which leads to the formal expansion
\begin{equation}
{\cal{G}}_{0}[{\cal{A}}]=G-G{\cal{A}}_{\mu 5}\Gamma_{\mu 5}G+\cdot\cdot\cdot.
\end{equation}
Here only the first-order dependence of ${\cal{G}}_{0}[{\cal{A}}]$ upon ${\cal{A}}_{\mu 5}$ is of interest and this will generate the nonperturbative axial vector vertex.

Substituting Eq.(14) and (16) into Eq.(13), we have the inhomogeneous ladder Bethe-Salpeter equation(BSE) of axial vector vertex, which reads
\begin{eqnarray}
&&\Gamma_{\mu 5}(y_1,y_2;z)\nonumber \\
&&=i\gamma_{\mu}\gamma_{5}\delta(y_1-y_2)\delta(y_1-z)
-\frac{4}{3}g^2_{s}D(y_1-y_2)\int du_{1}du_{2}\gamma_{\nu}G(y_1,u_1)\Gamma_{\mu 5}(u_1,u_2;z)G(u_1,y_2)\gamma_{\nu}
\end{eqnarray}

Fourier transform of Eq.(17) leads then to the momentum space form of the inhomogeneous BSE
\begin{equation}
\Gamma_{\mu 5}(P,q)
=i\gamma_{\mu}\gamma_{5}
-\frac{4}{3}\int\frac{d^4 K}{(2\pi)^4}g^2_{s}D(P-K)\gamma_{\nu}G(K+\frac{q}{2})\Gamma_{\mu 5}(K,q)G(K-\frac{q}{2})\gamma_{\nu}.
\end{equation}
As was shown above, both the rainbow DSE(10) and the ladder BSE(18) can be consistently derived from the action of the GCM in an external axial vector field ${\cal{A}}_{\mu 5}$.

From Eq.(18) it is evident that the $\Gamma_{\mu 5}(P,q)$, in addition to its Lorentz axial four-vector structure, is a four-by-four matrix in Dirac indices and is in general a function of two nonorthogonal four vectors, $P$ and $q$($P\cdot q\neq0$).

From Lorentz structure, the solution to $\Gamma_{\mu 5}$ must have the general form which fulfills Eq.(18), reads
\begin{equation}
\Gamma_{\mu 5}(P,q)=\gamma_{5}\Lambda^{(1)}_{\mu}(P,q)+i\gamma_{5}\gamma_{\nu}\Lambda^{(2)}_{\nu\mu}(P,q)+i\gamma_{\nu}\varepsilon_{\mu\nu\alpha\beta}p_{\alpha}q_{\beta}\eta^{-2}\Lambda^{(3)}(P,q).
\end{equation}
From which $\sigma_{\mu\nu}$ is absent, this is because the gluon two point function $g^{2}_{s}D(x-y)$ is limited to diagonal components in Lorentz indices. Further,the scalar matrix $1_{D}$ can be excluded by considering that the corresponding term has the form $1_{D}\Lambda_{\mu 5}$, where the matrix structure has been completely factored. In order to contribute to the axial vector vertex, this term must form a four axial vector, which requires $\Lambda_{\mu 5}$ to be an axial vector. It is immediately evident that this is impossible since there is an insufficient number of vectors($P$ and $q$) to combine with the tensor $\varepsilon_{\mu\nu\alpha\beta}$ to form an axial vector. The matrix structure in Eq.(19) is now explicit and the quantities $\Lambda^{(i)}$ are dimensionless function with Lorentz structure indicated by their indices. The constant $\eta$ has dimession of mass.

The further reduction of the $\Lambda^{(i)}$ to a set of invariant functions is achieved through the use of the symmetry transformation of the vertex $\Gamma_{\mu 5}(P,q)$ under $\gamma_{5}$, charge conjugation $C=\gamma_{2}\gamma_{4}$. The transformation properties are determined directly from the Eq.(18) and are given by

\begin{center}
\[
\left\{
\begin{array}{ll}
\gamma_{5}\Gamma_{\mu 5}(P,q)\gamma_{5} & =-\Gamma_{\mu 5}(-P,-q), \\
C\Gamma_{\mu 5}C^{-1} &=\Gamma^{t}_{\mu 5}(-P,q),
\end{array}.
\right.
\]
\end{center}
where $t$ denotes a matrix transpose. From the general form given above and Eq.(19), it is clear that $\Lambda^{(2)}_{\nu\mu}(P,q)$ and $\Lambda^{(3)}(P,q)$ are even in both $P$ and $q$, while $\Lambda^{(1)}_{\mu}(P,q)$ is odd in $q$ and even in $P$. The quantities $\Lambda^{(i)}$ can therefore be written as
\begin{center}
\[
\left\{
\begin{array}{ll}
\Lambda^{(1)}_{\mu}(P,q) & =\frac{q_{\mu}}{\eta}\lambda^{L}_{1}+\frac{P^{T}_{\mu}}{\eta}\frac{P\cdot q}{q^2}\lambda^{T}_{1}\\
\Lambda^{(2)}_{\nu\mu}(P,q) & =\frac{P_{\nu}q_{\mu}}{\eta^2}\frac{P\cdot q}{q^2}\lambda^{L}_{2}+\frac{P_{\nu}p^{T}_{\mu}}{\eta^2}\lambda^{T}_{2}
-\frac{q_{\nu}q_{\mu}}{q^2}\lambda^{L}_{3}-\left(\delta_{\nu\mu}-\frac{q_{\nu}q_{\mu}}{q^2}\right)\lambda^{T}_{3}+\frac{q_{\nu}P^{T}_{\mu}P\cdot q}{\eta^4}\lambda^{T}_{4}\\
\Lambda^{(3)}(P,q) & =\lambda^{T}_{5}.
\end{array}
\right.
\]
\end{center}
The eight scalar dimensionless coefficients $\lambda^{L}_{i}(i=1,2,3)$ and
$\lambda^{T}_{i}(i=1,2,3,4,5)$ depend on $P^2, q^2$ and $C^2_{Pq}$, where $C_{Pq}=P\cdot q/q^2$ is the direction cosine between $P$ and $q$. $p^{T}_{\mu}\equiv P_{\mu}-P\cdot q q_{\mu}/q^2$ is the vector transverse to $q_{\mu}$, i.e., $q_{\mu}p^{T}_{\mu}=0$. The advantage of the above decomposition lies in the fact that the longitudinal components $\lambda^{L}_{i}(i=1,2,3)$ are determined automatically from the quark propagator $G$ by means of the axial vector Ward-Takahashi identity(WTI),
\begin{equation}
q_{\mu}\Gamma_{\mu 5}(P,q)=G^{-1}(P+\frac{q}{2})\gamma_{5}+
\gamma_{5}G^{-1}(P-\frac{q}{2}),
\end{equation}
which can be easily verified by substituting it into Eq.(18) .

By means of the axial vector WTI(20) and Eq.(19), we have the longitudinal components($\lambda^{L}_{i}(i=1,2,3)$)
\begin{center}
\[
\left\{
\begin{array}{ll}
\lambda^{L}_{1}  & =\frac{\eta}{q^2}\left[B(P^2_{+})+B(P^2_{-})\right],\\
\lambda^{L}_{2} & =\frac{\eta^2}{P\cdot q}\left[A(P^2_{-})-A(P^2_{+})\right],\\
\lambda^{L}_{3}&=\frac{1}{2}\left[A(P^2_{+})+A(P^2_{-})\right],
\end{array}
\right.
\]
\end{center}
where $P_{\pm}\equiv P\pm q/2$.

This leaves only the five independent transversal components $\lambda^{T}_{i}(i=1,2,3,4,5)$ to be determined as solutions of inhomogeneous Bethe-Salpeter(BSE)(18). In order to do this, one has to project out the single $\lambda^{T}_{i}$ for each $i$ which can be done by multiplying Eq.(19) with appropriate Dirac matrices and taking traces. Performing the same operations on the right-hand side of Eq.(18) finally leads to a set of five couplied inhomogeneous integral equations for $\lambda^{T}_{i}$:
\begin{center}
\[
\left\{
\begin{array}{ll}
\frac{P^{T}_{\mu}}{4\eta}tr[\gamma_{5}\Gamma_{\mu 5}(P,q)] & = \frac{(P^{T})^2}{\eta^2}\frac{P\cdot q}{q^2} \lambda^{T}_{1}, \\
-\frac{i}{4}(\delta_{\mu\nu}-\frac{q_{\mu}q_{\nu}}{q^2})tr[\gamma_{\nu}\gamma_{5}\Gamma_{\mu 5}(P,q)] & = \frac{(P^{T})^2}{\eta^2}\lambda^{T}_{2}-3\lambda^{T}_{3},\\
-\frac{i}{4}\frac{P^{T}_{\mu}P^{T}_{\nu}}{(P^{T})^2}tr[\gamma_{\nu}\gamma_{5}\Gamma_{\mu}(P,q)]& = \frac{(P^{T})^2}{\eta^2}\lambda^{T}_{2}-\lambda^{T}_{3},\\
-\frac{i}{4}\frac{P^{T}_{\mu}q_{\nu}}{(P^{T})^2}tr[\gamma_{\nu}\gamma_{5}\Gamma_{\mu}(P,q)]& = \frac{P\cdot q}{\eta^2}\lambda^{T}_{2}+\frac{q^2 P\cdot q}{\eta^4}\lambda^{T}_{4},\\
tr[\gamma_{\nu}\Gamma_{\mu 5}(P,q)] & = 4i\epsilon_{\mu\nu\alpha\beta}P_{\alpha}q_{\beta}\eta^{-2}\lambda^{T}_{5}.
\end{array}
\right.
\]
\end{center}
The left-hand side of the above equation are evaluated using the inhomogenuous BSE for $\Gamma_{\mu 5}$(18). Based on the above five coupled integral equation and Eq.(18), in principle by means of numerical studies, we can get the the nonperturbative axial vector vertex $\Gamma_{\mu 5}$, which is useful for the calculation of nucleon spin in constituent quark model. So far, we have shown how the axial vector current of current quarks is related to that of constituent quarks within the framework of the global color symmetry model.

In order to estibalish qualitative analysis of the nonperturbative aspect of axial vector vertex, a simple model of the gluon two-point function is used as follow:
\begin{equation}
g^2 D((P-K)^2)=3\eta^2\pi^4\delta^{(4)}(P-K)
\end{equation}

Then Eq.(10) may be solved to give
\[
B(P^2)=(\eta^2-4 P^2)^{\frac{1}{2}},~~~~~~~A(P^2)=2 ~~~~for~~~~ P^2\leq\frac{\eta^2}{4},
\]
and
\[
B(P^2)=0,~~~~A(P^2)=\frac{1}{2}\left[1+(1+\frac{2\eta^2}{P^2})^{\frac{1}{2}}
\right]
~~~for~~~~ P^2\geq\frac{\eta^2}{4}.
\]
Here the scale parameter $\eta$ is a measure of the strength of the infrared slavery effect.

Using Eqs.(18) and (21), the above couplied integral equation can be further reduced to the following five couplied algebra equation.
\begin{eqnarray}
&&\frac{P\cdot q}{q^2}\lambda^{T}_{1}-\left[\alpha(P^2_{+})\beta(P^2_{-})+\alpha(P^2_{-})\beta(P^2_{+})\right]\eta^3\lambda^{T}_{3}\nonumber\\
&=&-\eta\left\{-\left[(P^2-\frac{q^2}{4})\alpha(P^2_{+})\alpha(P^2_{-})-\beta(P^2_{+})\beta(P^2_{-})\right]\eta\frac{P\cdot q}{q^2}\lambda^{T}_{1}
\right.\nonumber\\
&&+\left.\left[(P^2+\frac{P\cdot q}{2})\alpha(P^2_{+})\beta(P^2_{-})
+(P^2-\frac{P\cdot q}{2})\alpha(P^2_{-})\beta(P^2_{+})\right]\lambda^{T}_{2}\right.\nonumber\\
&&\left.+\left[(P\cdot q+\frac{q^2}{2})\alpha(P^2_{+})\beta(P^2_{-})+(P\cdot q-\frac{q^2}{2})\alpha(P^2_{-})\beta(P^2_{+})\right]\frac{P\cdot q}{\eta^2}\lambda^{T}_{4}\right\},
\end{eqnarray}
\begin{equation}
\lambda^{T}_{3}\left[1-\frac{\eta^2}{4}(2P^2-\frac{q^2}{2})\alpha(P^2_{+})\alpha(P^2_{-})-\frac{\eta^2}{2}\beta(P^2_{+})\beta(P^2_{-})\right]=1-\frac{q^2}{2}(P^T)^2\alpha(P^2_{+})\alpha(P^2_{-})\lambda^{T}_{5},
\end{equation}
\begin{equation}
\lambda^{T}_{5}\left[1+\frac{\eta^2}{2}\beta(P^2_{+})\beta(P^2_{-})-\frac{\eta^2}{2}(P^2-\frac{q^2}{4})\alpha(P^2_{+})\alpha(P^2_{-})\right]=\frac{\eta^4}{2}\alpha(P^2_{+})\alpha(P^2_{-})\lambda^{T}_{3},
\end{equation}
\begin{eqnarray}
&&\frac{P\cdot q}{\eta^2}\lambda^{T}_{2}+q^2\frac{P\cdot q}{\eta^4}\lambda^{T}_{4}\nonumber\\
&=&\frac{\eta^2}{2}\left\{-\frac{1}{\eta}\frac{P\cdot q}{q^2}\lambda^{T}_{1}\left[(P\cdot q+\frac{q^2}{2})\alpha(P^2_{+})\beta(P^2_{-})+(P\cdot q-\frac{q^2}{2})\alpha(P^2_{-})\beta(P^2_{+})\right]\right.\nonumber\\
&&\left.-\left[P\cdot q(P^2-\frac{q^2}{4})\right]\frac{\lambda^{T}_{2}}{\eta^2}
\alpha(P^2_{+})\alpha(P^2_{-})+\left[\frac{P\cdot q}{\eta^2}\lambda^{T}_{2}+q^2\frac{P\cdot q}{\eta^4}\lambda^{T}_{4}\right]\beta(P^2_{+})\beta(P^2_{-})\right. \nonumber\\
&&\left.+2P\cdot q\lambda^{T}_{3}\alpha(P^2_{+})\alpha(P^2_{-})-\left[2(P\cdot q)^2-q^2P^2-\frac{q^4}{4}\right]\frac{P\cdot q}{\eta^4}\lambda^{T}_{4}\alpha(P^2_{+})\alpha(P^2_{-})\right\},
\end{eqnarray}
\begin{eqnarray}
\lambda^{T}_{2}&=&\frac{\eta^4}{2}\left\{-\frac{\lambda^{T}_{1}}{\eta}\left[\alpha(P^2_{+})\beta(P^2_{-})+\alpha(P^2_{-})\beta(P^2_{+})\right]+\frac{1}{\eta^2}\lambda^{T}_{2}\beta(P^2_{+})\beta(P^2_{-})\right.\nonumber\\
&&\left.+\left[2\lambda^{T}_{3}-\frac{2}{\eta^4}(P\cdot q)^2\lambda^{T}_{4}-\frac{\lambda^{T}_{2}}{\eta^2}(P^2+\frac{q^2}{4})\right]\alpha(P^2_{+})\alpha(P^2_{-})-\frac{q^2}{\eta^2}\lambda^{T}_{5}\alpha(P^2_{+})\alpha(P^2_{-})\right\}.
\end{eqnarray}

For the region $P^2_{\pm}\leq\frac{\eta^2}{4}$, the analytic solution for the Eqs.(22-26) have the simple form
\[
\lambda^{T}_{1}=\frac{q^2}{P\cdot q}\left[\frac{4\eta[B(P^2_+)+B(P^2_-)]}{\eta^2-4P^2+5q^2+B(P^2_+)B(P^2_-)}\right],~~~~~~\lambda^{T}_{2}=-\frac{q^2}{\eta^2}\lambda^{T}_{4},
\]
\begin{equation}
\lambda^{T}_{3}=2\frac{[2\eta^2-4P^2+q^2+B(P^2_+)B(P^2_-)]}{3\eta^2-8P^2+6q^2},
\end{equation}
\[
\lambda^T_3=-\frac{32\eta^4}{[\eta^2-4P^2+5q^2+B(P^2_+)B(P^2_-)][3\eta^2-8P^2+6q^2]},~~~~~\lambda^T_5=\frac{8\eta^2}{3\eta^2-8P^2+6q^2},
\]
from which the singularity structure is visible. It appears from Eq.(27) that there is a pole of $q^2=-\eta^2/2$ for the region $P^2=C_{Pq}=0$.

To summarize, we have studied the nonperturbative aspect of axial vector vertex. This employes a consistent treatment of dressed quark propagator $G$ and the dressed axial vector vertex $\Gamma_{\mu 5}$, which are both determined from the model quark-quark interaction by the rainbow DSE for $G$ and the inhomogeneous ladder BSE for $\Gamma_{\mu 5}$.

\vspace*{4.cm}

\noindent{\large \bf Acknowledgement}

This work was supported in part by the National Natural Science Foundation
of China under Grant Nos 19975062, 10175033 and 10135030. 

\vspace*{2.cm}
\noindent{\large \bf References}
\begin{description}
\item{[1]} C. G. Callan, Jr., R. Dashen and D, J. Gross, Phys. Rev. {\bf D17}, 2717 (1978).
\item{[2]} E.~V. Shuryak, Nucl. Phys. {\bf B203}, 93 (1982); Phys. Rep. {\bf 115}, 151 (1984); T. Schafer andE. V. Shuryak. Rev. Mod. Phys. {\bf 70}, 323 (1998).
\item{[3]} D.~I. Dyakonov and X. Yu Petrov, Nucl. Phys. {\bf B245}, 259 (1984); {\bf B272}, 457 (1986).
\item{[4]} R. T. Cahill and C. D. Roberts, Phys. Rev. {\bf D32}, 2419 (1985).
\item{[5]} M. R. Frank and T. Meissner, Phys. Rev {\bf C53}, 2410 (1996); M. R. Frank and C. D. Roberts, Phys. Rev {\bf C53}, 390 (1996).
\item{[6]} C. D. Roberts and A. G. Williams, Prog. Part. Nucl. Phys.{\bf 33},
477 (1994), P. C. Tandy, Prog. Part. Nucl. Phys. 39, 117 (1997) and references therein.
\item{[7]} Xiao-fu L\"{u}, Yu-xin Liu, Hong-shi Zong and En-guang Zhao,
Phys. Rev. {\bf C58}, 1195 (1998), Hong-shi Zong, Xiao-fu L\"{u}, Jian-zhong Gu, Chao-hsi Chang and En-guang Zhao,Phys. Rev. {\bf C60}, 055208 (1999), Hong-shi Zong, Xiao-hua Wu, Xiao-fu L\"{u}, Chao-hsi Chang and En-guang Zhao,hep-ph/0109112.
\item{[8]} M. R. Frank, Phys. Rev. {\bf C51}, 987 (1995).
\item{[9]} M. R. Frank and P. C. Tandy, Phys. Rev. {\bf C49}, 478 (1994).
\item{[10]} T. Meissner and L. S. Kisslinger,  Phys. Rev. {\bf C59}, 986 (1999).
\end{description}

\end{document}